\documentclass[aps,prl,showpacs,amsmath,amssymb]{revtex4}

\usepackage{graphicx}% Include figure files
\usepackage{dcolumn}% Align table columns on decimal point
\usepackage{bm}% bold math

\newcommand{\be}{\begin{equation}}
\newcommand{\ee}{\end{equation}}
\newcommand{\bea}{\begin{eqnarray}}
\newcommand{\eea}{\end{eqnarray}}
\newcommand{\hf}{\frac12}
\newcommand{\nonu}{\nonumber\\}
\def\eq#1{(\ref{#1})}

\def\ord#1{{\cal O}\left(#1\right)}

\def\mr#1{{\mathrm#1}}

\begin{document}

\title{Running coupling constants of the Luttinger liquid}
\author{D. Boos\'e$^a$}\email{boose@lpt1.u-strasbg.fr}
\author{J. L. Jacquot$^a$}\email{jacquot@lpt1.u-strasbg.fr}
\author{J. Polonyi$^{a,b}$}\email{polonyi@fresnel.u-strasbg.fr}
\affiliation{$^a$Theoretical Physics Laboratory, CNRS and
Louis Pasteur University, Strasbourg, France}
\homepage{http://lpt1.u-strasbg.fr}
\affiliation{$^b$Department of Atomic Physics,
L. E\"otv\"os University, Budapest, Hungary}

\date{\today}

\begin{abstract}
We compute the one-loop expressions of two running coupling constants 
of the Luttinger model. The obtained
expressions have a nontrivial momentum dependence with Landau poles. The reason  
for the discrepancy between our results and those of other studies, which find 
that the scaling laws are trivial, is explained.
\end{abstract}

\pacs{71.10.Pm, 11.10.Hi, 73.22.Lp}% PACS, the Physics and Astronomy
\maketitle

\section{Introduction}
The Luttinger liquid, namely the one-dimensional system of 
interacting massless fermions with independent conservation laws for the 
left- and right-moving species, has been and still is the subject of 
many theoretical studies in   
Condensed Matter Physics (see, e.g., \cite{hald,lut,lutn}). This is 
due to the fact that it has several characteristic properties 
which are very different from those of a Fermi liquid.
Indeed, its momentum distribution 
is continuous at the Fermi energy even at zero temperature. Moreover, there
are no fermionic quasi-particles in the system because its single-particle
density of states vanishes as a power law when one approaches the 
Fermi energy. In addition, the Luttinger liquid has collective 
bosonic charge- and spin-density modes. The equations of motion 
of the system are solved with the help of the method of
bosonization \cite{bos}. This method allows to replace the 
complicated fermionic dynamics by another one which involves 
noninteracting bosonic degrees of freedom only.
The beta functions associated with the fermionic dynamics are believed to be equal to 
zero \cite{soly,mdc,shankar}, a result which is commonly
regarded as reflecting the triviality of the bosonic dynamics.   

The dynamics of the Luttinger liquid may actually be less trivial than 
expected. It may indeed well be that the particles in the system     
interact also by means of long range forces, as suggested by the following 
line of argument. It is known that the one-dimensional propagator of a free
massless fermion has a large anomalous dimension \cite{mltmsol,lsol}. This large  
anomalous dimension modifies the singularity structure
in momentum space in such a way that the scattering amplitude for isolated
particles vanishes together with the residue of the propagator
on the mass shell \cite{mltmsol,lsol}. One may therefore
say that a given particle moves slower or faster because of the
nature, repulsive or attractive, of the interaction rather than because
of the relative remoteness or closeness of other particles in the
system. Such a behaviour clearly suggests that the
interactions between massless fermions are able to act over large
distances in one dimension. The scale parameter controlling this behaviour can only
originate from a well-defined cutoff because the Luttinger model is known to
be classically scale invariant \cite{scala}.

This line of argument, however, doesn't provide a proof for the existence of long
range forces in the Luttinger liquid. Our aim in the present paper is to establish
in a more reliable way whether such forces exist
in the system or not. This is done with the help of
a renormalization group study of two effective coupling constants of the
model. The computations are done in the regularized version of the model with a
sharp cutoff. We find that the one-loop expressions of the running
coupling constants have Landau poles which line up in a scale invariant manner
around the Fermi points and at definite momentum scales
resulting from the finiteness of the cutoff. The
fact that the momentum dependence of these quantities is singular around
the Fermi points allows us to
conclude that long range forces do exist in the Luttinger liquid and so that the dynamics of this system is strongly coupled.

\section{Vertex Functions}
The regularized  Luttinger model with a sharp cutoff of value $\Lambda$ is defined by    
the action $S_\mr{Reg}(\Lambda)=S_0 +S_i(\Lambda)$, where
\bea\label{free}
S_0=\int_{{\bf {k}}}[\psi^*_-({\bf k})k_-\psi_-({\bf k}) + \psi^*_+({\bf k})k_+\psi_+({\bf k})]
\eea
describes the free propagation of massless fermions and
\bea\label{interaction}
S_i(\Lambda)=-g\int_{{\bf {k}},{\bf {k'}},{\bf {q}}}
\psi^*_+({\bf k}+{\bf q}) \psi_+({\bf k})\psi^*_-({\bf k'}-{\bf q})\psi_-({\bf k'})
\rho_+(\Lambda,k+q)\rho_+(\Lambda,k)
\rho_-(\Lambda,k'-q)\rho_-(\Lambda,k')
\eea
their interactions. This action is the simplest one
of all those considered in the context of the so-called g-ology models \cite{gology}.
The left- and right-moving spinless particles are represented by the
fields ($\psi_-$, $\psi^*_-$) and ($\psi_+$, $\psi^*_+$) respectively. We
use the conventions $\int_{{\bf k}}=\int dk_0 dk/(2 \pi)^2$ and 
${\bf k}=(k_0,k)$, where $k_0$ and $k$ are the frequency and 
momentum variables respectively. The momenta $k_\pm$ in Eq. \eq{free}
are defined by the relation $k_\pm=\pm{k}v_F-\mu-ik_0$, where 
$v_F$ is the Fermi velocity and $\mu={k_F}v_F$ the Fermi 
energy ($k_F$ is the Fermi momentum). The cutoff functions 
$\rho_{\pm}$ in Eq. \eq{interaction} are Heaviside functions $\Theta$; more
precisely, $\rho_{\pm}(\Lambda,k)=\Theta (\Lambda -|{k}\mp{k_F}|)$. We
assume in this paper that the absolute value of the bare coupling constant $g$
is small. It is important to bear in mind the fact that the expression of $S_0$, since
it contains the linearized form of the dispersion relation, is valid for small values
of $\Lambda$ only; the limit $\Lambda\to\infty$ is therefore a purely formal limit
in the context of the Luttinger model. We use here the method of the equation of motion
to compute the expressions of the vertex functions needed in the renormalization group
study. Although these expressions could be obtained by a straightforward application of
the Feynman rules, we prefer to follow the more compact computational scheme offered by
the chosen method because it shows in a more transparent manner the effect of the
regularization on the dynamics of the model.

Let us first recall that the generating functional $W[\eta^*_\pm,\eta_\pm]$ for the
connected $2n$-point Green functions $G^{(2n)}$ $(n=1,2,...)$ of the model is defined
by
\bea\label{functional}
e^W=\int D[\psi^*_-]D[\psi_-]D[\psi^*_+]D[\psi_+]
e^{-S_\mr{Reg}(\Lambda)-S_\mr{Src}},
\eea
with
\be
S_\mr{Src}=\int_{{\bf k}}\sum_{\alpha=\pm}[\eta^*_\alpha({\bf k})\psi_\alpha({\bf k})
+\psi^*_\alpha({\bf k})\eta_\alpha({\bf k})].
\ee
One has then $({\alpha_i=\pm}$, $i=1,...,n)$
\be
G^{(2n)}_{\alpha_1 \cdots \alpha_n} ({\bf p_1},\cdots,{\bf p_n},{\bf p'_1}, \cdots,{\bf p'_n})
=(2\pi)^{4n-2}\frac{\delta^{2n}W}{\delta\eta^*_{\alpha_1}({\bf p_1})\cdots\delta\eta^*_{\alpha_n}({\bf p_n})
\delta\eta_{\alpha_1}({\bf p'_1})\cdots\delta \eta_{\alpha_n}({\bf p'_n})}
\ee
Each Green function is regularized by the product of the cutoff functions corresponding to the 
external and internal lines. However, since all external momenta belong
necessarily to the intervals $[\pm{k_F}-\Lambda,\pm{k_F}+\Lambda]$,
the cutoff functions corresponding to the external lines can be safely ignored. 
Let us also recall that the Legendre transform $\Gamma[\psi^*_\pm,\psi_\pm]$ of $W[\eta^*_\pm,\eta_\pm]$ 
is the generating functional for the $2n$-point vertex functions $\Gamma^{(2n)}$ $(n=1,2,...)$ of 
the model. Indeed, one has $({\alpha_i=\pm}$, $i=1,...,n)$
\be
\Gamma^{(2n)}_{\alpha_1\cdots\alpha_n}({\bf p_1},\cdots,{\bf p_n},{\bf p'_1},\cdots,{\bf p'_n})
=(2\pi)^{4n-2}\frac{\delta^{2n}\Gamma}{\delta\psi^*_{\alpha_1}({\bf p_1})\cdots
\delta \psi^*_{\alpha_n}({\bf p_n})\delta\psi_{\alpha_1}({\bf p'_1})\cdots
\delta\psi_{\alpha_n}({\bf p'_n})}.
\ee
The two effective coupling constants considered in this paper are those related   
to the four-point vertex functions $\Gamma_{+-}^{(4)}$
and $\Gamma_{++}^{(4)}$, whose expressions are 
computed below in the one-loop approximation.

Infinitesimal translations $\psi_\alpha\to\psi_\alpha+\epsilon$ $({\alpha=\pm})$                   
of the integration variables $\psi_\alpha$
in Eq. \eq{functional} lead to the two regularized
equations of motion of the model, one for each species of particles. These two  
equations may be combined into a single one, namely,
\bea\label{equationmotion1}
0&=&\biggl\{(2\pi)^2p_\alpha\frac{\delta}{\delta\eta_\alpha({\bf p})}+\eta^*_\alpha({\bf p})\nonu
&&+g(2\pi)^6\rho_\alpha(\Lambda,p)\int_{{\bf k},{\bf q}}
\biggl[\delta_{\alpha,+} \biggl(\rho_+(\Lambda,p+q)\rho_-(\Lambda,k-q)\rho_-(\Lambda,k)\frac{\delta^3}
{\delta\eta_+({\bf p}+{\bf q})\delta\eta_-({\bf k}-{\bf q})\delta\eta_-^*({\bf k})} \biggr)\nonu
&&+\delta_{\alpha,-} \biggl(\rho_-(\Lambda,p+q)\rho_+(\Lambda,k-q)\rho_+(\Lambda,k)\frac{\delta^3}                                                                               
{\delta\eta_-({\bf p}+{\bf q})\delta\eta_+({\bf k}-{\bf q})\delta\eta_+^*({\bf k})} \biggr)\biggr]\biggr\}e^W.
\eea
If we let the functional derivative $\delta/\delta\eta^*_+({\bf p'})$ 
act on the regularized equation of motion corresponding to the case $\alpha=+$, we 
obtain the following regularized Dyson-Schwinger equation:
\bea\label{equationmotion2}
0&=&p_+G^{(2)}_+({\bf p'},{\bf p})+\delta({\bf p}-{\bf p'})+g\int_{{\bf k},{\bf q}}\biggl[G^{(4)}_{+-}({\bf p'},{\bf k},{\bf p}+{\bf q},{\bf k}-{\bf q})\nonu
&&-(2\pi)^2G^{(2)}_-({\bf k},{\bf k}-{\bf q}) G^{(2)}_+({\bf p'},{\bf p}+{\bf q})\biggr]
\rho_+(\Lambda,p+q)\rho_-(\Lambda,k-q)
\rho_-(\Lambda,k).
\eea

The action of well-chosen third-order functional derivatives on the 
regularized equation of motion corresponding to the case $\alpha=+$, Eq. \eq{equationmotion1}, leads
to the identities needed to compute the one-loop expressions of the vertex functions 
$\Gamma_{+-}^{(4)}$ and $\Gamma_{++}^{(4)}$. If we let the functional derivative
$\delta^3/\delta\eta^*_+({\bf p'})\delta\eta_-({\bf r})\delta\eta^*_-({\bf r'})$ act on this equation 
of motion and then use the 
regularized Dyson-Schwinger equation, Eq. \eq{equationmotion2}, we obtain the following identity:
\bea\label{equationvertex2}
0&=&p_+G^{(4)}_{-+}({\bf r'},{\bf p'},{\bf r},{\bf p})-g\int_{{\bf k},{\bf q}}\biggl[         
G^{(6)}_{--+}({\bf r'},{\bf k},{\bf p'},{\bf r},{\bf k}-{\bf q},{\bf p}+{\bf q})+(2\pi)^2 \biggl(G^{(2)}_+({\bf p'},{\bf p}+{\bf q})
G^{(4)}_{--}({\bf r'},{\bf k},{\bf r},{\bf k}-{\bf q})\nonu
&&-G^{(2)}_-({\bf r'},{\bf k}-{\bf q})  G^{(4)}_{+-}({\bf p'},{\bf k},{\bf p}+{\bf q},{\bf r})+G^{(2)}_-({\bf k},{\bf k}-{\bf q})
G^{(4)}_{-+}({\bf r'},{\bf p'},{\bf r},{\bf p+q})
-G^{(2)}_-({\bf k},{\bf r})G^{(4)}_{-+}({\bf r'},{\bf p'},{\bf k}-{\bf q},{\bf p}+{\bf q}) \biggr)\nonu
&&+(2\pi)^4G^{(2)}_{-}({\bf k},{\bf r})G^{(2)}_{-}({\bf r'},{\bf k}-{\bf q})G^{(2)}_{+}({\bf p'},{\bf p}+{\bf q})\biggr]             
\rho_+(\Lambda,p+q)\rho_-(\Lambda,k-q)\rho_-(\Lambda,k).
\eea
The one-loop expression of $\Gamma_{+-}^{(4)}$ is obtained from this  
identity in the following way. First, we express each connected four- or six-point 
Green function in Eq. \eq{equationvertex2}
in terms of its associated vertex function. Next, we replace each   
vertex function $\Gamma^{(4)}_{\alpha_1\alpha_2}({\bf p},{\bf r},{\bf p'},{\bf r'})$ (${\alpha_1}\neq{\alpha_2}$) 
that has appeared in the integrand of the equation 
by its expression at leading order in $g$, namely, 
\bea\label{leading}
\Gamma^{(4)}_{\alpha_1\alpha_2}({\bf p},{\bf r},{\bf p'},{\bf r'})
= g\delta({\bf p}+{\bf r}-{\bf p'}-{\bf r'})(\delta_{\alpha_1,+}\delta_{\alpha_2,-}              
+\delta_{\alpha_1,-}\delta_{\alpha_2,+})+\ord{g^2}           
\eea
(the vertex functions $\Gamma^{(4)}_{--}$ and $\Gamma^{(6)}_{--+}$ need not be 
considered because their expansion in powers of the bare coupling constant
starts with a power of $g$ larger than one). In a last step, we simplify the  
integrand by using the expression of the regularized Dyson-Schwinger equation 
at leading order in $g$, namely,
\be
\frac{1}{G^{(2)}_+({\bf p})}=-p_++g\int_{{\bf k}}G^{(2)}_-({\bf k})\rho_-(\Lambda,k)+\ord{g^2}.
\ee
We find that 
\bea\label{ghgam}
\Gamma^{(4)}_{+-}({\bf p},{\bf r},{\bf p'})=g+g^2\int_{{\bf q}}
\biggl[\frac{\rho_+(\Lambda,p'+q)}{(p'+q)_+}
\frac{\rho_-(\Lambda,r+q)}{(r+q)_-}
+\frac{\rho_-(\Lambda,p+r-p'-q)}{(p+r-p'-q)_-}
\frac{\rho_+(\Lambda,p'+q)}{(p'+q)_+}\biggr]+\ord{g^3},
\eea
where $\Gamma^{(4)}_{+-}({\bf p},{\bf r},{\bf p'})$ is introduced by means of the relation
$\Gamma^{(4)}_{+-}({\bf p},{\bf r},{\bf p'},{\bf r'})=\delta ({\bf p}+{\bf r}-{\bf p'}-{\bf r'})\Gamma^{(4)}_{+-}({\bf p},{\bf r},{\bf p'})$.

If we let the functional derivative
$\delta^3/\delta\eta^*_+({\bf r'})\delta\eta_+({\bf r})\delta\eta^*_+({\bf p'})$ act 
on the regularized equation of motion corresponding to the case $\alpha=+$, Eq. \eq{equationmotion1}, and
then use the regularized Dyson-Schwinger equation, Eq. \eq{equationmotion2}, we obtain another identity
which is the following:
\bea\label{equationvertex3}
0&=&p_+G^{(4)}_{++}({\bf r'},{\bf p'},{\bf r},{\bf p})-g\int_{{\bf k},{\bf q}}\biggl[         
G^{(6)}_{+-+}({\bf r'},{\bf k},{\bf p'},{\bf r},{\bf k}-{\bf q},{\bf p}+{\bf q})+(2\pi)^2 \biggl(G^{(2)}_+({\bf p'},{\bf p}+{\bf q})
G^{(4)}_{+-}({\bf r'},{\bf k},{\bf r},{\bf k}-{\bf q})\\
&&-G^{(2)}_+({\bf r'},{\bf p}+{\bf q})  G^{(4)}_{+-}({\bf p'},{\bf k},{\bf r},{\bf k}-{\bf q})+G^{(2)}_-({\bf k},{\bf k}-{\bf q})
G^{(4)}_{++}({\bf r'},{\bf p'},{\bf r},{\bf p}+{\bf q}) \biggr) \biggr] 
\rho_+(\Lambda,p+q)  \rho_-(\Lambda,k-q)\rho_-(\Lambda,k).\nonumber
\eea
The one-loop expression of $\Gamma^{(4)}_{++}$ is obtained
from this identity by following the same procedure
as for $\Gamma^{(4)}_{+-}$. We find that    
\bea\label{result}
\Gamma^{(4)}_{++}({\bf p},{\bf r},{\bf p'})=g^2\int_{{\bf q}}
\biggl[\frac{\rho_-(\Lambda,q)}{q_-}
\frac{\rho_-(\Lambda,q+r-p')}{(q+r-p')_-}
-\frac{\rho_-(\Lambda,q)}{q_-}
\frac{\rho_-(\Lambda,q+p-p')}{(q+p-p')_-}\biggr]+\ord{g^3},
\eea
where $\Gamma^{(4)}_{++}({\bf p},{\bf r},{\bf p'})$ is introduced by means of the relation
$\Gamma^{(4)}_{++}({\bf p},{\bf r},{\bf p'},{\bf r'})=\delta ({\bf p}+{\bf r}-{\bf p'}-{\bf r'})\Gamma^{(4)}_{++}({\bf p},{\bf r},{\bf p'})$.

\section{Running coupling constants}
For later convenience, the variables of the vertex functions
$\Gamma^{(4)}_{+\sigma}$ ($\sigma=\pm$) are parametrized as
\bea
{\bf k}&=(-iv_F(k+\Delta_k),k_F+k),~~~
{\bf l}&=(-iv_F(l+\Delta_l),\sigma(k_F+l)),\nonu
{\bf k'}&=(-iv_F(k'+\Delta_{k'}),k_F+k'),~~~
{\bf l'}&=(-iv_F(l'+\Delta_{l'}),\sigma(k_F+l')),
\eea
where the momentum variables $\Delta_k$,
$\Delta_l$, $\Delta_{k'}$, and $\Delta_{l'}$ measure
the distance from the mass shell. Conservation of energy imposes the condition
$k+\Delta_k+l+\Delta_l=k'+\Delta_{k'}+l'+\Delta_{l'}$
and conservation of momentum the condition $k$+$\sigma l$=$k'$+$\sigma l'$. The
combination of these two conditions leads to the additional constraint
$\Delta_k+\Delta_l=\Delta_{k'}+\Delta_{l'}$ in the case of the
vertex function $\Gamma^{(4)}_{++}$. The mass shell expression of $\Gamma^{(4)}_{+-}$
depends on two momenta only (say, $k$ and $l$) because the combination of the
two conservation conditions with the condition
$\Delta_k=\Delta_l=\Delta_{k'}=\Delta_{l'}=0$ gives the
additional constraints $k=k'$ and $l=l'$ in the case of this
vertex function.
Performing the integrations in Eqs. \eq{ghgam} and \eq{result}, we find that
\bea\label{gamn}
\Gamma^{(4)}_{+-}({\bf k},{\bf l},{\bf k}')
&=&g+\frac{g^2}{4\pi v_F}f_{+-}({\bf k},{\bf l},{\bf k}')+\ord{g^3},\nonu
\Gamma^{(4)}_{++}({\bf k},{\bf l},{\bf k}')
&=&\frac{g^2}{4\pi v_F}f_{++}({\bf k},{\bf l},{\bf k}')+\ord{g^3},
\eea
where
\bea\label{equationvertex6}
f_{+-}({\bf k},{\bf l},{\bf k}')&=&\ln\left[\frac{(k'+l)^2
-(k'-l+\Delta_{k'}-\Delta_l)^2}
{(k-l)^2-(k+l+\Delta_k+\Delta_l)^2}\right]
+\ln\left[\frac{(2\Lambda-|k-l|)^2
-(k+l+\Delta_k+\Delta_l)^2}
{(2\Lambda-|k'+l|)^2
-(k'-l+\Delta_{k'}-\Delta_l)^2}\right],\nonu
f_{++}({\bf k},{\bf l},{\bf k}')&=&\frac{k-k'}
{k-k'+\hf(\Delta_k-\Delta_{k'})}
-\frac{k'-l}{k'-l+\hf(\Delta_{k'}-\Delta_{l})}.
\eea

The renormalization group method allows us, starting from the truncated expressions of the vertex
functions, Eqs. \eq{gamn}, \eq{equationvertex6}, to compute the expressions of the running coupling constants
we are interested in. This method is usually implemented in three steps
which, in the present context, are the following.

(i) In a first step, we introduce two effective coupling constants $\tilde g_{+-}$ and $\tilde g_{++}$
by choosing a subtraction point $(\tilde{\bf k},\tilde{\bf l},\tilde{\bf k}')$ in the  
six-dimensional energy-momentum space and setting 
\bea\label{effect}
\tilde g_{+-}&=&Re\Gamma^{(4)}_{+-}(\tilde{\bf k},\tilde{\bf l},\tilde{\bf k}'),\nonu 
\tilde g_{++}&=&\Gamma^{(4)}_{++}(\tilde{\bf k},\tilde{\bf l},\tilde{\bf k}').
\eea
The subtraction point is chosen in such a manner that the absolute
value of each effective coupling constant is
small. It has to be noticed that if $\tilde g_{+-}$ is already present in the bare
action, $\tilde g_{++}$ is fully generated by the dynamics.

(ii) Next, we choose a scheme, a trajectory in the energy-momentum
space which can be parametrized by a variable, denoted
by $\lambda$ in the sequel.
The trajectory crosses the  subtraction point for $\lambda=\tilde\lambda$.
The trajectory usually chosen is the one associated with
an homogeneous rescaling of the variables $k$,
$\Delta_k$, etc., in which case $\lambda$ controls the
scale dependence.

(iii) The values of the effective coupling constants $g_{+\sigma}(\lambda)$ ($\sigma=\pm$) corresponding
to an arbitrary point of the trajectory, that is, to a given value of
$\lambda$, are then computed by integrating the beta functions along this trajectory
from the subtraction point up to the considered point. The
beta functions $\beta_\sigma(\lambda)$ give the rate of variation of the effective coupling constants
as $\lambda$ increases, namely,
\be\label{nontrbfu}
\beta_\sigma(\lambda)=\lambda\frac{dg_{+\sigma}(\lambda)}{d\lambda}
=\frac{g_{+\sigma}^2{(\lambda)}}{4\pi v_F}\lambda\frac{df_{+\sigma}(\lambda)}{d\lambda}+\ord{g_{+\sigma}^3}.
\ee
The integration of these quantities leads to the following expression of the running coupling constants:
\be\label{trajint}
g_{+\sigma}(\lambda)=\frac{\tilde g_{+\sigma}}{1+\frac{\tilde g_{+\sigma}}{4\pi v_F}
[f_{+\sigma}(\tilde\lambda)-f_{+\sigma}(\lambda)]},
\ee
with $\tilde g_{+\sigma}$=$g_{+\sigma}(\tilde\lambda)$. This expression, which gives at once
the values of the effective coupling constants at any point of the trajectory, may be
viewed as resulting from an optimized partial resummation of the full perturbation series of the
vertex functions.

Notice that the particular trajectory which defines the scheme does not appear in the
expression of Eq. \eq{trajint}. This circumstance allows us to generalize this
expression to an arbitrary scheme, namely,
\be\label{gensol}
g_{+\sigma}({\bf k},{\bf l},{\bf k}')=\frac{\tilde g_{+\sigma}}
{1+\frac{\tilde g_{+\sigma}}{4\pi v_F}[f_{+\sigma}(\tilde{\bf k},\tilde{\bf l},\tilde{\bf k}')
-f_{+\sigma}({\bf k},{\bf l},{\bf k}')]}.
\ee
This expression offers the advantage that the effective coupling constants depend
only on the coordinates $({\bf k},{\bf l},{\bf k}')$ of the point at which they are computed. Its
derivation is based upon the fact that the rate of change of the renormalized coupling constants
caused by an infinitesimal step in the energy-momentum space has a component along each direction
of this space. The expression of such a component is
\be\label{genbfc}
\beta_{q,\sigma}(\lambda)=q\frac{\partial g_{+\sigma}(q)}{\partial q}
=\frac{g_{+\sigma}^2{(q)}}{4\pi v_F}q
\frac{\partial f_{+\sigma}(q)}{\partial q}+\ord{g_{+\sigma}^3},
\ee
where the symbol $q$ represents a component of
$({\bf k},{\bf l},{\bf k}')$. The
integration of these "directional" beta functions along the chosen trajectory leads
to the expression of Eq. \eq{gensol} which, interestingly enough, may also be viewed as originating from
an optimized partial resummation of the Bethe-Salpeter equation.

A mass shell expression of $g_{+-}$ which is
valid in the particle-particle ($p$-$p$) channel is obtained by
using Eq. \eq{trajint} in the case of the linear trajectory
$k(\lambda)=l(\lambda)=\lambda\Lambda$ ($|\lambda|<1$). We find that  
\be\label{pppm}
g^{p-p}_{+-}(\lambda)=\frac{g^{p-p}_{+-}(\tilde\lambda)}{1+
\frac{g^{p-p}_{+-}(\tilde\lambda)}{4\pi v_F}\left[
\ln\left(\frac{1+|\tilde\lambda|}{1-|\tilde\lambda|}\right)
-\ln\left(\frac{1+|\lambda|}{1-|\lambda|}\right)\right]}.
\ee
This equation clearly diverges and the Landau pole is found at
\be
\lambda^{p-p}_L=\frac{1+\tilde\lambda-(1-\tilde\lambda)
e^{-\frac{4\pi v_F}{g^{p-p}_{+-}(\tilde\lambda)}}}
{1+\tilde\lambda+(1-\tilde\lambda)
e^{-\frac{4\pi v_F}{g^{p-p}_{+-}(\tilde\lambda)}}}.
\ee
The effective interaction between the particles has obviously not the same 
sign in the particle-hole channel as in the particle-particle channel. It is 
easy to verify that the function $f_{+-}$, Eq. \eq{equationvertex6}, the one 
used to define the effective coupling constant $\tilde g_{+-}$, Eq. \eq{effect}, changes 
sign if the momentum $l$ is tranformed into $-l$; a mass shell expression
of $g_{+-}$ which is valid in the particle-hole ($p$-$h$) channel can therefore be obtained by
using Eq. \eq{trajint} in the case of the linear trajectory
$k(\lambda)=-l(\lambda)=\lambda\Lambda$ ($|\lambda|<1$). The result is 
\be\label{phpm}
g^{p-h}_{+-}(\lambda)=\frac{g^{p-h}_{+-}(\tilde\lambda)}{1-
\frac{g^{p-h}_{+-}(\tilde\lambda)}{4\pi v_F}\left[
\ln\left(\frac{1+|\tilde\lambda|}{1-|\tilde\lambda|}\right)
-\ln\left(\frac{1+|\lambda|}{1-|\lambda|}\right)\right]}.
\ee
This equation diverges too and the Landau pole is at
\be
\lambda^{p-h}_L=\frac{1+\tilde\lambda-(1-\tilde\lambda)
e^{+\frac{4\pi v_F}{g^{p-h}_{+-}(\tilde\lambda)}}}
{1+\tilde\lambda+(1-\tilde\lambda)
e^{+\frac{4\pi v_F}{g^{p-h}_{+-}(\tilde\lambda)}}}.
\ee

We use now Eq. \eq{gensol} to obtain $g_{+-}$ on the mass shell,
\be\label{pmin}
g_{+-}(k,l)=\frac{\tilde g_{+-}}
{1+\frac{\tilde g_{+-}}{4\pi v_F}\left[
\ln\left[\frac{(2\Lambda-|\widetilde{k}-\widetilde{l}|)^2
-(\widetilde{k}+\widetilde{l})^2}{(2\Lambda-|\widetilde{k}+\widetilde{l}|)^2
-(\widetilde{k}-\widetilde{l})^2}\right]
-\ln\left[\frac{(2\Lambda-|k-l|)^2-(k+l)^2}
{(2\Lambda-|k+l|)^2-(k-l)^2}\right]\right]}.
\ee
Notice that the first term on the right-hand side of the first equation
in Eq. \eq{equationvertex6} is vanishing on the mass shell.
As a result, the scale dependence, i.e. the deviation
from a scale invariant result, arises from the presence
of the cutoff $\Lambda$ only. The Landau poles
corresponding to the divergences of this expression
belong to the line of the plane $(k,l)$,
\be\label{genlppm}
\frac{4\pi v_F}{\tilde g_{+-}}=\ln\left[\frac{(2\Lambda-|k-l|)^2
-(k+l)^2}{(2\Lambda-|k+l|)^2-(k-l)^2}\right]
-\ln\left[\frac{(2\Lambda-|\widetilde{k}-\widetilde{l}|)^2
-(\widetilde{k}+\widetilde{l})^2}{(2\Lambda-|\widetilde{k}+\widetilde{l}|)^2
-(\widetilde{k}-\widetilde{l})^2}\right].
\ee
This line stays away from the Fermi points
for any generic choice of the bare coupling constant, that is, in
the absence of fine tuning. The behaviour of $g_{+-}$
that is valid slightly away from the mass shell is
similar to the one of $g_{+-}$  on the mass shell
and the corresponding equations can be obtained from
Eqs. \eq{pppm}-\eq{genlppm} by adding terms of the order $\ord{\Delta}$ on the
right-hand sides.

The coupling strength $g_{++}$ is always vanishing at the order of one loop
on mass shell. Using Eq. \eq{gensol} again, we obtain
\be\label{pplus}
g_{++}=\frac{\tilde g_{++}}
{1+\frac{\tilde g_{++}}{4\pi v_F}\left[\frac{\widetilde{k}-\widetilde{k'}}
{\widetilde{k}-\widetilde{k'}+\hf(\widetilde\Delta_k-\widetilde\Delta_{k'})}
-\frac{\widetilde{k'}-\widetilde{l}}{\widetilde{k'}-\widetilde{l}
+\hf(\widetilde\Delta_{k'}-\widetilde\Delta_l)}
-\frac{k-k'}{k-k'+\hf(\Delta_k-\Delta_{k'})}
+\frac{k'-l}{k'-l+\hf(\Delta_{k'}-\Delta_l)}\right]}
\ee
off-shell. The Landau poles are now spread over a five-dimensional hypersurface
whose implicit equation is
\be\label{genlppp}
\frac{4\pi v_F}{\tilde g_{++}}
=\frac{k-k'}{k-k'+\hf(\Delta_k-\Delta_{k'})}
-\frac{k'-l}{k'-l+\hf(\Delta_{k'}-\Delta_l)}
-\frac{\widetilde{k}-\widetilde{k'}}{\widetilde{k}-\widetilde{k'}+\hf(\widetilde\Delta_k-\widetilde\Delta_{k'})}
+\frac{\widetilde{k'}-\widetilde{l}}{\widetilde{k'}-\widetilde{l}+\hf(\widetilde\Delta_{k'}-\widetilde\Delta_l)}.
\ee
This hypersurface approaches the Fermi point of the right-moving particles
in the limit $\Delta_k,\Delta_l$, and $\Delta_{k'}\to0$ for any value
of $\tilde g_{++}$.

Several comments are in order at this point.
We start with the scale dependence of the coupling constants
$g_{+-}$ and $g_{++}$. The latter is scale invariant according to Eq. \eq{pplus}
and its beta function is vanishing in the conventional schemes where the
variables $k$, $\Delta_k$, etc. are scaled homogeneously.
The scale invariance of $g_{+-}$ is broken by the cutoff, as shown by
Eq. \eq{pmin}. The limit $\Lambda\to\infty$ carried out in a
given finite order of the perturbation expansion suppresses this
breakdown and one would naively expect that the renormalized
theory recovers the classical scale invariance. But the partial infinite resummation
of the perturbation expansion, carried out by the generalized
renormalization group scheme, indicates that the
interactions become strong
within certain kinematical regions because of
the accumulation of large contributions
of higher orders in the perturbation expansion.
These regions are characterized by the scale of
the cutoff. This prevents us from removing the cutoff
and restoring the naive, classical scale invariance
at the same time.

Another remark is about the nature of the singularities of the effective
coupling strengths. The divergence in Eqs. \eq{pmin} and
\eq{pplus}
does not necesserily mean that the theory becomes strongly coupled.
In fact, the effective interaction strengths should be defined by
scattering amplitudes or other physical observables which reflect the interactions
among elementary or dressed excitations. Such a construction always
involves a finite resolution in the energy-momentum space which can be
imagined as coming from a smearing of some Green functions
in the kinematical space. As long as the singularities are integrable,
as is the case of $g_{+-}$ in Eq. \eq{pmin},
such a smeared coupling strength remains finite despite the diverging
numerical values the effective coupling constant may take in the
vicinity of the Landau poles. But the non-integrable
singularity of $g_{++}$ appearing in Eq. \eq{pplus}
produces a diverging coupling strength
for an arbitrary localisation of the constituents
in the energy-momentum space.

We close this Section by pointing out the way the apparent contradiction
between our results and the well-known vanishing of the usual beta function can be
resolved.

(i) The beta function computed in Refs. \cite{soly,mdc} is based upon a
particular choice of the subtraction point, namely
$p_0=r_0=p_0'=r_0'=0$ ($\Delta_k=-k, \Delta_l=-l, \Delta_{k'}=-k', \Delta_l'=-l'$)
and $k=-l=-k'=-l'/3$. The special feature of this point is that
the radiative corrections $f_{+-}$ and $f_{++}$, given by Eqs.
\eq{equationvertex6}, are vanishing.
This subtraction point together with the usual, homogeneous
scaling implies vanishing beta functions. But the generalized scheme
used in this work picks up the radiative corrections arising from an
arbitrary displacement in the kinematical space and reveals the non-naturalness
of this cancellation.

(ii) Another argument, the one used in Ref. \cite{shankar}, rests on the
study of the behaviour of the bare coupling constant in the blocking
procedure. The bare coupling constant was defined in that work at the
Fermi point and the possible energy-momentum dependence was treated
perturbatively. The point is that the leading order result, the
$\ord{g^2}$ coupling constant
at the Fermi point, agrees with Eq. \eq{ghgam}. The singular energy-momentum
dependence generated by the partial resummation in Eqs. \eq{pmin} and
\eq{pplus} suggests that the dependence of the coupling constant on
the energy-momentum can not be taken into account by expanding
around the Fermi point and the simplicity of the scenario spelled out
in Ref. \cite{shankar} appears misleading.

(iii) A more formal argument has also been used to justify the triviality
of the scaling laws. This argument is based upon the (supposed) equivalence between
the behaviour of the running coupling constant in
two different limits, the one where the cutoff is
removed and the one where each energy-momentum variable
$\bf k$ tends towards zero. Indeed, since
these variables and $\Lambda$ are the only scales
($k_F$ is absent from the beta functions),
the dimensionless running coupling
constants should not depend on them separately
but rather on the ratios ${\bf k}/\Lambda$. Consequently, the
limit ${\bf k}\to0$ is supposed to coincide with the limit
$\Lambda\to\infty$ where the scale dependence
drops, as one can see by inspecting Eqs.
\eq{equationvertex6}. The fallacy of this argument is usually shown
by recalling the phenomenon of dimension transmutation \cite{dimtr}.
This phenomenon is found in asymptotically free, classically scale invariant
models where the cutoff generates a scale which remains finite even
in a renormalized theory where the cutoff is already removed.
The Luttinger model offers a simpler mechanism to break
the scale invariance by its Landau poles, as pointed out after Eq.
\eq{genlppp} above.

\section{Summary and conclusion}
Our aim in this paper was to establish the existence of long range forces
in the Luttinger liquid in a reliable manner. This has been done with the
help of a renormalization group study of two effective coupling constants
of the model. The computations have been done in the one-loop order.
We have shown that the running coupling constants have Landau poles
in the vicinity of the Fermi points for arbitrarily weak bare interactions.
The singular momentum dependence of the effective coupling constants in the
vicinity of the Fermi points shows that the interactions
between the fermionic excitations are of long range.

The problem of the breakdown of the perturbation expansion in the vicinity of the
Fermi points is difficult to solve.
It is known that the correspondence between the fermionic and bosonic versions
of the model can be established at any finite order
of the perturbation expansion in powers of the particle mass or interaction
strength. However, since the strong interactions close to the Fermi
points modify the single particle dispersion relation in
a fundamental manner, the perturbative approach
may not be sufficient to prove this correspondence.
We believe that a more systematic study of the scaling laws
established in both versions of the theory is
needed to clarify this question.

The partial resummation, performed by the integration of the beta
functions, remains consistent as long as the effective coupling strengths
are weak. When the one-loop beta functions predict
large values for the coupling constants,
the computation is no longer reliable and one
usually assumes that the true dynamics is indeed strongly coupled
in terms of the degrees of freedom used. A similar problem,
namely the one of the soft gluons, blocks the way
towards the analytical understanding
of Hadronic Physics. We believe that any progress
towards more definite conclusions about the dynamics of the Luttinger model
requires some non-perturbative method in order to treat the
strong, long range interactions, just as in the case of
QCD.

\begin{acknowledgments}
We wish to acknowledge useful discussions with D. Cabra and
J. S\'olyom and to thank K. Kikoin for
bringing Ref. \cite{lsol} to our attention as well as the Referee 
for helping us to improve the content of the paper.
\end{acknowledgments}

\end{document}